\documentclass[prd,superscriptaddress,onecolumn,showkeys]{revtex4}
\usepackage{textcomp}
\usepackage{eurosym}
\usepackage{amsfonts}
\usepackage{array}
\usepackage{amsthm}
\usepackage{bm}
\usepackage{palatino}
\usepackage[colorinlistoftodos]{todonotes}
\usepackage{mathpazo}
\usepackage{supertabular}
\usepackage{subfig}
\usepackage{amssymb}
\usepackage{eurosym}
\usepackage{amsmath}
\usepackage{epsfig}
\usepackage{float}
\usepackage{graphics}
\usepackage{changes}
\usepackage{color}
\usepackage{graphicx}
\usepackage[colorlinks=true,
            linkcolor=blue,
           urlcolor=blue,
           citecolor=blue]{hyperref}

\def\be{\begin{equation}}
\def\ee{\end{equation}}
\def\beq{\begin{eqnarray}}
\def\eeq{\end{eqnarray}}

\def\bes{\begin{eqnarray}}
\def\ees{\end{eqnarray}}

\newcommand\myeq{\mathrel{\overset{\makebox[0pt]{\mbox{\normalfont\tiny\sffamily $e\rightarrow\infty$}}}{=}}}
\begin{document}
\title{Exploring light deflection and black hole shadows in Rastall theory with plasma effects}

\author{Riasat Ali}
%\orcidlink{0000-0002-9371-1113}
\email{riasatyasin@gmail.com}
\affiliation{Department of Mathematics, Shanghai University and Newtouch Center for Mathematics of Shanghai University, Shanghai-200444, People's Republic of China}

\author{Xia Tiecheng}
%\orcidlink{0000-0001-8235-0319}
\email{xiatc@shu.edu.cn}
\affiliation{Department of Mathematics, Shanghai University and Newtouch Center for Mathematics of Shanghai University, Shanghai-200444, People's Republic of China}

\author{Rimsha Babar}
%\orcidlink{0000-0002-2650-7796}
\email{rimsha.babar10@gmail.com}
\affiliation{Department of Mathematics, GC
University Faisalabad Layyah Campus, Layyah-31200, Pakistan}

\author{Ali \"Ovg\"un}
\email{ali.ovgun@emu.edu.tr}
%\orcidlink{0000-0002-9889-342X}
\affiliation{Physics Department, Eastern Mediterranean University, Famagusta, 99628 North Cyprus via Mersin 10, Turkiye}

\date{\today}

\begin{abstract}
In this article, we examine the gravitational deflection of particles in curved spacetime immersed in perfect fluid in the context of Rastall theory. We propose an infinite region approach to Gibbons-Werner to avoid singularity, given that the integral region is generally infinite. In the Rastall theory framework, the black hole solutions in the dust field are studied. Additionally, we check the deflection angle from this spacetime under the influence of plasma. Furthermore, we analytically compute plasma's impact on a black hole shadow using a ray-tracing approach and Hamiltonian equation. Hence, the light ray motion equations are independent of the plasma's velocity. It is assumed that plasma is a dispersive medium, pressureless and non-magnetised, and the plasma particle density corresponds to particle accumulation. The supermassive black hole's shadow and emitted energy are explored when plasma falls radially from infinity onto the black hole.
\end{abstract}


\date{\today}

\maketitle

\section{Introduction}

In general relativity (GR), one of the most significant fields of study is the movement of objects in intense gravitational fields. An object's path can be examined using the geodesic expression of the intense gravitational field. By considering it as a test particle, its gravitational field is ignorable. An essential consideration in observing the test particle's velocity is the trajectory's deflection. There are two types of particles: massive particles and massless particles, mainly photons. Gravitational theories \cite{R1} such as GR are verified in significant part for massless particles through gravitational deflection. Many techniques have been suggested \cite{R2}-\cite{R22} to study the deflection angle (DA) of photons. For massive particles, they may function as representatives of all that exists. Later on, Eddington's detection justified \cite{R23} the deflection of light moving by the sun. Gravitational lensing has been extensively studied as a powerful tool in various fields of cosmology and astronomy.
For neutrinos, gravitons, neutrons, and celestial events of high energy ($\pi$-mesons, $K$-mesons, $\mu$ons, etc.) are examples. There are also theoretically weakly associating massive particles and their axions \cite{R24}. Essential details as to the source, lens, trajectory context, and particles can be obtained from the gravitational deflection of these significant particles \cite{R25, R26, R27, R28, R29, R30, R31, R32, R33, R34, R35, R36}. The Gibbons-Werner (GW) approach, initially proposed by Gibbons-Werner in $2008$ \cite{R37} and then modified by scientists in more recent times \cite{R38, R39, R40, R41,  R43, R44}, helps in the geometric explanation and determination of the DA for both massless and massive particles. The development of a four-dimensional spacetime that can be utilized to explain the motion and integral region of the particle on a two-dimensional Riemann manifold, the trajectory of the particle, an auxiliary circular curve, a radial upward curve that passes through the source, and a radial outward curve that passes through the observer are all part of the fundamental technique of the GW approach. The Gauss-Bonnet theorem (GBT) can be applied to the integral region for the DA in the form of geometric parameters.

Consider a black hole (BH), which has a consistent spacetime structure but can be lighted by external electromagnetic radiation with varying shapes, colors, and behaviors throughout time. Many mathematical approaches can detect the BH, but its impact on light distribution can be observed. The Event Horizon Telescope Collaboration \cite{R45} appears as the first image of a BH. The photo shows a continuous spacetime structure illuminated by a time-varying emission zone, encouraging more investigation and explanation. Several concepts employed by the Event Horizon Telescope (EHT), such as BH shadows, have been given multiple explanations in the literature. In \cite{R46}, Synge calculated the angular radius of the Schwarzschild BH's shadow using a static observer model. The investigation has been done on the various geometries of BH shadows in Refs \cite{R47} -\cite{R76}. As the magnetic field parameter grows, the shadows of the Schwarzschild and Kerr BHs immersed in the Melvin magnetic field grow and elongate horizontally. Black hole shadows can only be predicted by employing a ray-tracing approach \cite{R57, R58, R62} for light motion systems with non-integrability.

However, the work has focused on the direct local effects of cosmic models on the known BH solutions. Babichev et al. \cite{R77} demonstrate that in a scenario with a phantom field, the accreting particles of the phantom scalar field into the central BH cause the BH mass to decrease. But this does have a broad influence. A modified metric that includes the BHs surrounding spacetime can be used to determine the local changes in the spacetime geometry surrounding the fundamental BH. In this context, Kiselev \cite{R78} has achieved an analytically static, spherically symmetric solution to Einstein's equations. The basis of the Rastall theory is related to high-curvature situations; hence, the physics of BHs may provide a suitable framework for further exploration of this theory. Thus, as a novel class of non-vacuum BH solutions to this theory, our study aims to find the surrounding Kiselev-like BH solutions. The properties of the BHs surrounding fields, typically such as radiation, dust, or a dark energy component, define this solution \cite{R78, R79}. The Kiselev every field solution's exceptional cases (a) Schwarzschild BH encompassed by quintessence, BH encompassed by magnetized Ernst field and a Reissner-Nordstrom BH encircled by radiation and dust (b), as well as its phase changes and thermodynamical quantities are examined in \cite{R79, R80} and \cite{R81} describes the dynamics of a neutral and a charged particle surrounding the Schwarzschild black encircled by quintessence matter.

One of the main reasons for researching the BH surrounding field in Rastall theory is to avoid spacetime singularity. Although Rastall's theory is well-developed, numerous attempts have been made, such as the black rings field theory and BH geometry. Almost all these approaches acknowledge the expectation of an intrinsic extended structure in spacetime (Gibbons-Werner approach). Stable-orbit photons constantly move around a BH; they cannot leave the BH or travel to infinity. In this instance, the pro-grade stable photon orbits correspond to the dark region that only partially emerges from the primary shadow. The BH shadow is half-panoramic (equatorial) due to the lack of backward unstable (stable) light rings. In this instance, the BH shadow turns into an equatorial, panoramic shadow without a grey area for stable photon orbits. The expectation that there should be an intrinsic extended structure in spacetime is acknowledged by almost all of these approaches (ray-tracing approach). It is studied that the BH shadows are immersed in charge fields and that the BH shadow is affected by the perfect fluid in the context of Rastall theory. 

This is the way the paper is formatted. Section \ref{sec2} discusses the analytical technique for spherically symmetric BHs in Rastall theory.
Section \ref{sec3} studies the DA for BHs occupied by dust fields and their graphical analysis. In section \ref{sec4}, we investigate the results of DA in the background of the plasma frame for associated BH. Section \ref{sec5} comprises the shadow cast and contour plots for BHs surrounded by dust fields. In section \ref{sec6}, we summarize the results of our work.

\section{Introductory review of a black hole in the background of Rastall theory}\label{sec2}

The structure of the energy-momentum tensor of the matter field has a non-minimal coupling; Rastall gravity was a basic mathematical extension of GR \cite{R81a}. The Rastall theory's first explanation was that the conservation law of local energy and momentum in a flat spacetime does not always imply that it will also be conserved in a curved spacetime in a matter field that includes Rastall gravity's new features for the gravitational theory.

In this Rastall theory of gravity framework, we seek the general non-vacuum spherically symmetric static uncharged BH solutions within the section. Utilizing Rastall's idea \cite{R82, R83, A1}, we obtain the following for a spacetime where an energy-momentum source of $T_{ab}$ occupies the Ricci scalar $R$ as
\begin{equation}
\lambda R^{,b}= T^{ab};a ,
\end{equation}
where $\lambda$ represents the Rastall parameter, the conventional GR conservation law that has deviated. The Rastall field equations can be expressed as
\begin{equation}
kT_{ab}=G_{ab}+k\lambda g_{ab}R,
\end{equation}
where the gravitational constant of Rastall coupling is denoted by $k$. In the limit of $\lambda \rightarrow 0 $ and $k=8\pi G$, these field equations reduce to GR field equations, with $G$ as the gravitational constant of Newton coupling.
To derive BH solutions, we take the usual Schwarzschild coordinates general spherical symmetric metric as  \cite{A1}
\begin{equation}
ds^{2}=-A(r)dt^{2}+\frac{dr^{2}}{A(r)}+
r^2d\Omega^{2},\label{st}
\end{equation}
with the two-dimensional unit sphere is represented by the equation $d\Omega^{2}=d\theta^{2}+\sin^{2}\theta d\phi^{2}$, and $f(r)$ is a fundamental metric function that depends on the radial coordinate. We can derive non-vanishing components of the Rastall tensor, defined as $H_{ab}=G_{ab}+k\lambda g_{ab}R$, which can be followed by applying this metric, we get  \cite{A1}
\begin{eqnarray}
H^{0}_{0} &=& G^{0}_{0}+k\lambda R=-\frac{1}{A}G_{00}+k\lambda R=\frac{1}{r^{2}}(\acute{A}r-1+A)+k\lambda R,\nonumber\\
H^{1}_{1} &=& G^{1}_{1}+k\lambda R=AG_{11}+k\lambda R=\frac{1}{r^{2}}(\acute{A}r-1+A)+k\lambda R,\nonumber\\ 
H^{2}_{2} &=& G^{2}_{2}+k\lambda R=-\frac{1}{r^{2}}G_{22}+k\lambda R=\frac{1}{r^{2}}(\acute{A}r+\frac{1}{2}r^{2}\acute{\acute{A}})+k\lambda R,\nonumber\\
H^{3}_{3} &=& G^{3}_{3}+k\lambda R=-\frac{1}{r^{2}\sin^{2}\theta}G_{33}+k\lambda R=\frac{1}{r^{2}}(\acute{A}r+\frac{1}{2}r^{2}\acute{\acute{A}})+k\lambda R. \label{RA1}
\end{eqnarray}
In this case, the Ricci scalar is defined as
\begin{equation}
R=-\frac{1}{r^{2}}(r^{2}\acute{\acute{A}}+4r\acute{A}+2+2A),
\end{equation}
with the derivative about the radial coordinate r is represented by the prime sign. In relation to the Rastall tensor ($H^{a}_{b}$) non-vanishing components, the following diagonal form should be present in the entire energy-momentum tensor that governs this spacetime.
\begin{equation}
T^{a}_{b}=\begin{pmatrix}
T^{0}_{0} & 0 & 0 & 0\\
0 & T^{1}_{1}& 0 & 0\\
0 & 0 & T^{2}_{2}& 0\\
0 & 0 & 0 & T^{3}_{3}
\end{pmatrix},
\end{equation}
with it is satisfied with the Rastall tensor $H^{a}_{b}$ symmetry features. For the solutions of Rastall tensor, the equality conditions $H^{0}_{0}=H^{1}_{1}$ and $H^{2}_{2}=H^{3}_{3}$ must also have $T^{0}_{0}=T^{1}_{1}$ and $T^{2}_{2}=T^{3}_{3}$, respectively. After that, a general energy and momentum tensor T with these symmetry features can be generated in the manner shown as
\begin{equation}
T^{a}_{b}=\tau^{a}_{b}+E^{a}_{b},\label{RA2}
\end{equation}
with $E^{a}_{b}$ is the trace free of Maxwell tensor defined by
\begin{equation}
E^{a}_{b}=\frac{2}{k}(F_{a\mu}F^{\mu}_{b}-\frac{1}{4}g_{ab}F^{\mu\nu}F_{{\mu\nu}}).\label{E1}
\end{equation}
So that $F_{ab}$ represents the anti-symmetric Faraday tensor obeying the corresponding vacuum Maxwell expression as
\begin{eqnarray}
F^{ab}_{;a} &=& 0,\nonumber\\  
\partial[\alpha F_{ab}] &=&0. \label{FT}
\end{eqnarray}
Given that the spacetime metric (\ref{st}) has spherical symmetry, the only non-vanishing Faraday tensor $F^{ab}$ of components is imposed to be $F^{01}=-F^{10}$. Next, using the equations in (\ref{FT}), one can get
\begin{equation}
F^{01}=\frac{Q}{r^{2}},\label{F1}
\end{equation}
where $Q$ represents an integration constant behaving as an electrostatic charge, the only Maxwell tensor $E^{a}_{b}$ non-vanishing components are given by equations (\ref{st}), (\ref{E1}), and (\ref{F1}) as
\begin{equation}
E^{a}_{b}=\frac{Q^{2}}{kr^{4}}\textit{diagonal}(-1, -1, 1, 1),\label{RA3}
\end{equation}
possessing the symmetries in the $H^{a}_{b}$ tensor and evidently expressing an electrostatic field. Conversely, $\tau^{a}_{b}$ denotes the surrounding field's energy-momentum tensor, which is defined \cite{R84} as
\begin{eqnarray}
\tau^{0}_{0} &=& -\rho_{d}(r)\nonumber\\
\tau^{j}_{i} &=& -\rho_{d}(r)\mu\big(-\frac{r_{j}r^{i}}{r_{n}r^{n}}-3\nu\frac{r_{j}r^{i}}{r_{n}r^{n}}+\nu \delta^{j}_{i}\big).\label{RA4}
\end{eqnarray}
For the arbitrary parameters $\mu$ and $\nu$, which correspond to the internal structure of the BH surrounding the field, this model of $\tau^{a}_{b}$ implies that the space component is proportionate to the time category, which represents the energy density. In this case, the surrounding field is indicated by the subscript (d), which is usually any combination of dust fields. We can determine the isotropic average over the angles by using \cite{R84} the
\begin{equation}
<\tau^{j}_{i}> = \frac{1}{3}\rho_{d}\delta^{j}_{i}=\rho_{d}\delta^{j}_{i},
\end{equation}
as it is assumed that $<r^{j}r_{i}> = \frac{\mu}{3}\rho_{d}\delta^{j}_{i}r_{n}r^{n}.$
The expression of equilibrium for the surrounding field becomes as
\begin{equation}
p_{d} = \rho_{d}\omega_{d},~~~~\omega_{d}=\frac{\mu}{3},
\end{equation}
with $\omega_{d}$ and $\rho_{d}$ indicate the equation of the state parameter and the pressure, respectively. The principle of the addition and linearity scenario assumed in reference \cite{R84} to find the free parameter $\nu$ of the energy-momentum tensor of the surrounding field as suggested is thus precisely provided by the field expressions (\ref{RA1}) for the entire energy-momentum tensor in (\ref{RA2}), (\ref{RA3}) and (\ref{RA4}) as
\begin{equation}
\nu =-\frac{3\omega_{d}+1}{6\omega_{d}}.
\end{equation}
Then, the following form can be used to derive the non-vanishing components of the $\tau_{ab}$ tensor 
\begin{eqnarray}
\tau^{0}_{0}&=&\tau^{1}_{1}=-\rho_{d}\nonumber\\
\tau^{2}_{2} &=& \tau^{3}_{3}=\frac{1}{2}(3\omega_{d}+1)\rho_{d},
\end{eqnarray}
which in the Rastall tensor $H^{a}_{b}$ also have identical symmetries. Consequently, all of $H^{a}_{b}$ symmetry is admitted by our whole constructed energy-momentum tensor in (\ref{RA2}). The Rastall field equations $\tau^{a}_{b}$ can be regarded as the only energy-momentum tensor of assistance. The solutions produced will thus explain the encircled uncharged BH solutions in the Rastall theory context, which are not the same as the ones in GR. Within the context of this theory, the most generic class of static surrounding charged BH solutions can be obtained by including the Maxwell tensor $E^{a}_{b}$ in $T^{a}_{b}$. The field equations are solved, and their general solution is obtained. Next, we tackle the two uncharged/charged solutions.
From the $H^{0}_{0}=T^{0}_{0}$ and $H^{1}_{1}=T^{1}_{1}$ components of the Rastall field expression, the differential equation that follows is obtained as
\begin{equation}
\frac{1}{r^{2}}(r\acute{A}+A-1)-\frac{k \lambda }{r^{2}}(r^{2}\acute{\acute{A}}+4r\acute{A}+2\acute{A}-2)=-\frac{Q^{2}}{r^{4}}-k\rho_{d}, \label{RA5}
\end{equation}
and $H^{2}_{2}=T^{2}_{2}$ and $H^{3}_{3}=T^{3}_{3}$
components are interpreted as
\begin{equation}
\frac{1}{r^{2}}(r\acute{A}+\frac{1}{2}r^{2}\acute{\acute{A}})-\frac{k \lambda }{r^{2}}(r^{2}\acute{\acute{A}}+4r\acute{A}+2A-2)=-\frac{Q^{2}}{r^{4}}+\frac{k\rho_{d}}{2}(1+3\omega_{d}),\label{RA6} 
\end{equation}
As a result \cite{A1}, the two differential equations (\ref{RA5}) and (\ref{RA6}) allow us to analytically compute  the one unknown functions $A(r)$ as
\begin{equation}
A(r)=1-\frac{2M}{r}+\frac{Q^{2}}{r^{2}}
-\frac{N_{d}}{r^{\frac{1-3\omega_d-6k\lambda(1+\omega_d)}{1-3k\lambda(1+\omega_d)}}},\label{x1}
\end{equation}
the BH mass and the surrounding field structure parameter are represented by two integration constants, $M$ and $N_d$, respectively. The parameters $k$ and $\lambda$ represent the Rastall geometric parameters, and $\omega_d$ denotes the equation of state parameter of the BH surrounding field. Remember that the surrounding field's characteristics are represented by the integration constant $N_d$. Any combination of $k$, $\lambda$, and $\omega_d$ parameters can accept various positive or negative $N_d$ values. We recover the Reissner-Nordstr\"{o}m BH occupied by a surrounding field in GR, which was initially discovered by Kiselev \cite{R84}, in the limit of $\lambda\rightarrow 0$ and $k=8\pi G N$. The metric in Eq. (\ref{x1}) is a new static solution with interesting features. Now, we will study the surrounding BH by the dust radiation, quintessence, cosmological constant, and phantom fields as the sub-classes of the general solution of Eq. (\ref{x1}) and their interesting aspects in detail. The two differential equations (\ref{RA5}) and (\ref{RA6}) allow us to analytically compute other one unknown function $\rho_{d}(r)$ as
\begin{equation}
\rho_{d}(r)=-\frac{3W_d N_d}{kr^{\frac{3(1+\omega_d)-12k\lambda(1+\omega_d)}{1-3k\lambda(1+\omega_d)}}},
\end{equation}
where the field structure parameter can be represented by
\begin{equation}
W_{d}(r)=-\frac{(1-4k\lambda)(k\lambda(1+\omega_d)-\omega_d)}{(1-3k\lambda(1+\omega_d))^2}.
\end{equation}
Remember that the surrounding field's characteristics are represented by the integration constant $N_d$. We have $\rho_d(r)=-\frac{3}{k}W_d N_d r^{-3(1+\omega_d)}$ where $W_d=\omega_d$ for $\lambda=0$, in the GR limit. The BH metric (\ref{x1}) surrounded by the dust field in Rastall theory can be presented as 
\begin{equation}
ds^{2}=A(r)dt^{2}+B(r)dr^{2}+
C(r)d\theta^{2}+D(r)d\phi^{2},\label{x2}
\end{equation}
where 
\begin{equation}
A(r)=\frac{1}{B(r)}=1-\frac{2M}{r}+\frac{Q^{2}}{r^{2}}
-\frac{N_{d}}{r^{\frac{1-6k\lambda}{1-3k\lambda}}},
~~~C(r)=r^{2},~~~D(r)=r^{2}\sin^{2}\theta,\label{x21}
\end{equation}
and $M$ is a mass of BH, $N_{d}$ is the dust field structure  parameter and $Q$ is a charge of BH. This metric is not the same as the metric of the charged BH in GR \cite{R84} that is encircled by a dust field. As one can see, the BH in the dust background in GR appears as a charged BH with an effective mass $M_{e}=2M+N_d$ in the limit of $\lambda\rightarrow 0$ and $k=8\pi G N$. We may observe that, for $k \lambda \neq 0$, the Rastall theory's geometric parameters $k$ and $\lambda$ can be crucial in determining different solutions concerning GR. For $k\lambda \neq 0$, the Rastall correction term gives the BH a distinct character incomparable to the mass and charge terms and never behaves like the mass or charge terms. Due to the Rastall geometric parameters, such nontrivial character might significantly alter the thermodynamics, causal structure, and Penrose diagrams compared to GR. For this case, the geometric parameter $W_d$ can be defined as
\begin{equation}
W_d=\frac{k\lambda(1-4k\lambda)}{(1-3k\lambda)^2}.\label{x3}
\end{equation}
Next, for the weak energy condition denoted by the relation $W_d N_d\leq 0$, we require $N_d > 0$ for $0\leq k\lambda<\frac{1}{4}$ and $N_d<0$ for the field structure constant for $k\lambda > \frac{1}{4}$.
In this instance, $W_d$ and $\rho_d$ are essentially distinct from their GR versions and the density $\rho_d$ can be given as 
\begin{equation}
\rho_d=\frac{3\lambda(1-4k\lambda)N_d}{(1-3k\lambda)^2}r^{-\frac{3-12k\lambda}{1-3k\lambda}}.\label{x4}
\end{equation}
Moreover, the geometry of the Rastall theory allows us to derive an effective equation of state parameter $\omega_{eff}$ for the modification term can be written as
\begin{equation}
\omega_{eff}=\frac{1}{3}\left(-1+\frac{1-6k\lambda}{1-3k\lambda}\right).\label{x5}
\end{equation}
It may be observed that, except for the $k\lambda=0$ corresponding to the GR limit, $\omega_{eff}$ can never be zero. Therefore, this theory's solutions essentially diverge from GR. The Rastall solution of BH thermodynamical quantities is examined in \cite{R85a, R85b}. Regarding Eq. (\ref{x5}), there are two distinct and intriguing groups. For the case, $\frac{1}{6}<k\lambda<\frac{1}{3}$ that implies to $\omega_{eff}\leq\frac{1}{3}$. Here, dark energy is represented by an adequate surrounding fluid with an effective equation of state parameter $\omega_e$, which results in an effective repulsive gravitational pull. This illustrates that, for a given $k$, the stronger the acceleration phase, or the more substantial coupling $g_{\mu\nu}R$ in Rastall theory, the larger values of $\lambda$. 
When $k\lambda<\frac{1}{6}\bigcup k\lambda>\frac{1}{3}$ implies to $\omega_{eff}\geq-\frac{1}{3}$ is the result. Here, we have an adequate surrounding fluid with the usual attractive gravitational effect and whose equation of state parameter respects the strong energy condition. Depending on the value of the effective equation of state parameter, this could lead to the universe expanding more slowly or even contracting.

\section{Light deflection in Rastall theory}\label{sec3}

This section is based on studying DA in a non-plasma frame for a BH surrounded by a dust field in Rastall theory. The generalized GW technique \cite{R85, R86} significantly improves the associated computation and provides a thorough framework for describing the GW approach for a wide range of scenarios. We think our study will implement the GW method in BH physics much more easily. When the GW approach is applied and developed for spacetimes, there is an ill-defined infinite integral region for some spacetimes expressing singular behavior, and the computation required is complex. In particular, we study that the radial coordinate of the additional circular arc can be chosen freely by carefully examining the integrals of geodesic curvature along the support circular arc and of Gaussian curvature throughout the integral region. As a result, the ill-defined condition is resolved, and the integral region without singular performance can be formed. With complex computation, we construct a simplified formula to determine the DA in a few steps. This formula is based on the free choice of the auxiliary circular arc and the reduction of the integral of geodesic curvature along the path. The efficiency and effectiveness of our approach are persuasively validated when we finally compute the DA of particles in Kiselev spacetime in Rastall gravity. Furthermore, we provide the DA of particles for the charged solution in Rastall gravity for the first time. The following expression can be used to find the DA for the Kiselev solution using GBT in the non-singular domain region. When we consider the null condition $ds^{2}=0$ at equatorial plane $(\theta =\frac{\pi}{2})$ for the metric (\ref{x2}) when the tropical area is where the source, observer, and null photon are all located then the optical metric can be obtained as
\begin{equation}
dt^{2}=\frac{B(r)}{A(r)}dr^{2}+\frac{r^{2}}{A(r)}d{\phi}.\label{x6}
\end{equation}
We rewrite the above metric in the form
\begin{equation}
dt^2=Y(r)dr^2+Z(r)d\phi,\label{xx6}
\end{equation}
where
\begin{equation}
Y(r)=\frac{B(r)}{A(r)},~~~~~Z=\frac{r^2}{A(r)}.
\end{equation}
To calculate the Ricci scalar for the given metric, at first, we compute the non-zero Christoffel symbols in the following manner
\begin{equation}
\Gamma^{1}_{11}=\frac{Z'(r)}{2Z(r)}-\frac{Y'(r)}{2Y(r)},~~~\Gamma^{2}_{12}=\frac{Y(r)}{r}-\frac{Y'(r)}{2},~~~\Gamma^{1}_{22}=\frac{r^{2} Y'(r)}{2Y(r)Z(r)}-\frac{r}{Z(r)},\label{x7}
\end{equation}
where $1$ and $2$, represent the $r$ and $\phi$ coordinates, respectively. The Ricci scalar is obtained as follows
\begin{equation}
R= \frac{2Z^{''}(r)Y(r)Z(r)-Z'(r)Y'(r)Z(r)-Z'^2(r)Y(r)}{2Y^2(r)Z^2(r)}.\label{x8}
\end{equation}
The Gaussian curvature can be derived by using the following formula
\begin{equation}
K=\frac{R}{2}.\label{xx9}
\end{equation}
Using Eq. (\ref{x9}), the Gaussian curvature for a BH surrounded by a dust field is calculated as 
\begin{eqnarray}
K &\approx &\frac{2M}{r^{3}}+\frac{(1+6k\lambda) N_d}{r^3}-\frac{3Q^{2}}{r^{4}}+\frac{3(2k\lambda-1)M N_d}{r^4}
+\frac{6 M Q^2}{r^{5}}\nonumber\\&+&\frac{3(2-\lambda)N_d Q^2}{2r^5}
+O(M^{2},Q^{3},\lambda^{2}).\label{x9}
\end{eqnarray}
The DA is determined by applying the optical Gaussian curvature to a non-singular region $W_e$, bounded by $\partial W_e = \omega_{\tilde{g}}\bigcup U_{e}$. Alternatively, a non-singular domain with the Euler characteristic $\alpha(W_e) = 1$ outside the light trajectory can be used. The GBT for this area can be expressed as \cite{R37}
\begin{equation}
    \int\int_{W_{e}}{K dS}+\oint_{\partial W_e}{\kappa dt}+\sum_{i}\epsilon_{i}
    =2\pi\alpha(W_e).\label{x10}
\end{equation}
As $k=\bar{g}(\nabla_{\dot{\varpi}}\dot{\varpi},\ddot{\varpi})$ and $\bar{g}(\dot{\varpi},\dot{\varpi})=1,~
\ddot{\varpi}$ imply unite acceleration vector, and $\epsilon_i$ expresses the exterior angle at the $ith$ vertex, $\kappa$ represents geodesic curvature in the above expression. The associated jump angles drop to $\frac{\pi}{2}$ as $e\rightarrow\infty$, yielding $\theta_{O}+\theta_{d}\rightarrow\pi$. Thus,
\begin{equation}
\int\int_{W_e}{K dS}+\oint_{\partial W_{e}}{\kappa dt}+\epsilon_{i}=2\pi\alpha(W_{e}),\label{x11}
\end{equation}
here, jump angle is represented by $\epsilon_{i}=\pi$.  Obtaining the geodesic curvature when $e\rightarrow\infty$ in the following way
\begin{equation}
\kappa(U_{e})=\mid\nabla\dot{_{U_{e}}}\dot{U_{e}}\mid.\label{x12}
\end{equation}
Since the radial component of geodesic curvature is computed as
\begin{equation}
(\nabla_{\dot{U_{e}}}\dot{U_{e}})^{r}=\dot{U^{\phi}_{e}}
\partial_{\phi}\dot{U_{e}^{r}}+\Gamma^{1}_{22}
(\dot{U}^{\phi}_{e})^{2}.\label{x13}
\end{equation}
For large value of $e, U_{e}:= r(\phi)=e=\text{constant}$, then the result is
\begin{equation}
\lim_{e\rightarrow \infty}\kappa(U_{e})=\lim_{e\rightarrow \infty}(\nabla_{\dot{U_{e}}}\dot{U_{e}})^{r}\rightarrow\frac{1}{e}\label{x14}.
\end{equation}
Since there is no topological defect in the geodesic curvature, $K(U_{e})\rightarrow e^{-1}$. However, by using the optical metric Eq. (\ref{x6}), it can be written as follows  
\begin{equation}
\lim_{e\rightarrow \infty}dt = e d\phi.\label{x15} 
\end{equation} 
Using Eq. (\ref{x14}) and (\ref{x15}), we obtain
\begin{equation}
K(U_{e})dt=d\phi.\label{x16}
\end{equation}
Now, using Eq. (\ref{x16}), we can get the following
equation
\begin{equation}
\int\int_{W_{e}}{K dS}+\oint_{\partial W_{e}}{\kappa dt}~ \myeq \int\int_{W\infty}{K dS}+
\int^{\pi+\psi}_{0}d\phi.\label{x17}
 \end{equation}
We employ the straight-line approximation, $r=\frac{b}{\sin \phi}$, where $b$ represents the impact parameter. The Gibbons and Werner approach uses the Gauss-Bonnet theorem to determine the DA as \cite{R37}
\begin{equation}
\psi\approx -\int^{\pi}_{0}\int^{\infty}_\frac{b}{sin\phi}K\sqrt{\det\vec{g}}dr d\phi,\label{x18}
\end{equation}
where $\sqrt{\det\vec{g}}$ is computed as
\begin{equation}
\sqrt{\det\vec{g}}=\frac{3M}{r^2}+\frac{1}{r^{3}}-\frac{3Q^2}{2r^{3}}+\frac{3M N_{d}}{2r^{1+\frac{1-6k\lambda}{1-3k\lambda}}}.\label{xx19}
\end{equation}
To compute the angle of deflection, use the Gaussian curvature up to the leading order terms and obtain
\begin{eqnarray}
\psi_{1} &\approx & \frac{4M}{b}-\frac{2N_d}{b}-\frac{12N_d k\lambda}{b}-\frac{3Q^2\pi}{4b^{2}}-\frac{3M N_d\pi}{2b^2}-\frac{3M N_d k\lambda\pi}{2b^2}-\frac{4NQ^2}{3b^3}\nonumber\\
&+&\frac{8MQ^{2}}{3b^{3}}+\frac{2N_dQ^2 k \lambda}{3b^3}+O(M^{2},Q^{4},\lambda^{2}).\label{x19}
\end{eqnarray}
The DA for a BH surrounded by a dust field depends on impact parameter $b$, BH mass $M$, charge $Q$, Rastall geometric parameters $k$, $\lambda$, and dust field structure parameter $N_d$. We can also observe that the impact parameter is inverse to the angle $\psi$. Moreover, it is worth mentioning here that, in the absence of structure parameter $N_d=0$ and Rastall parameters $k\lambda=0$, the above result reduces into DA of Reissner–Nordstr\"{o}m BH \cite{W1, W2} as well as in the absence of charge $Q=0$, we recover the DA of Schwarzschild BH $\psi_{Sch}=4M/b$.

\begin{figure}[H]
\centering
\includegraphics[width=5cm,height=5cm]{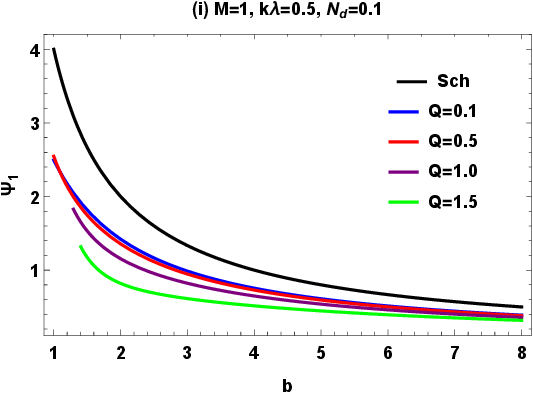}\includegraphics[width=5cm,height=5cm]{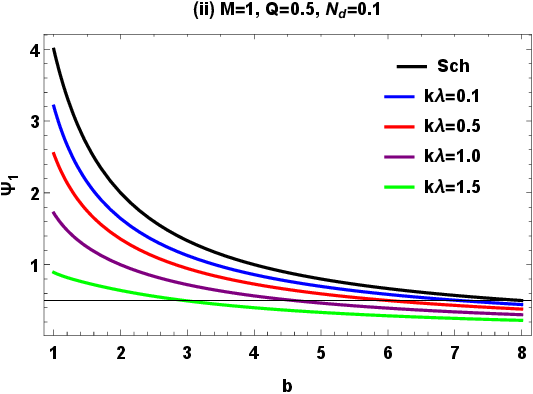}
\caption{Deflection angle $\psi_1$ versus impact parameter $b$ with variations of charge $Q$ and Rastall parameters $k \lambda$ with fixed mass $M=1$ and dust field structure parameter $N_d=0.1$}. \label{D1}
\end{figure}
In Fig. \ref{D1}: the left panel shows the variations of charge $Q$ with fixed mass $M=1$, Rastall parameters $k \lambda=0.5$ and dust field structure parameter $N_d=0.1$ and right panel represents the variations of Rastall parameters $k \lambda$ with fixed mass $M=1$, charge $Q=0.5$ and dust field structure parameter $N_d=0.1$. In both panels, the black curve gives the Schwarzschild solution. We can observe that the DA for rising values of charge and Rastall parameters from a BH surrounded by a dust field is decreasing via rising impact parameter, and it attains an asymptotically flat form till $b\rightarrow\infty$. A strong deflection can be observed for a small value of impact parameter $b=1.2$. Graphically, we can also see the inverse relationship between impact parameter $b$ and DA $\psi_1$.

\section{Light deflection in Rastall theory under the influence of non-magnetic plasma}\label{sec4}

A plasma and a dust field can affect the DA of light in the context of gravitational lensing. The plasma interacts directly with photons through its charged particles, causing additional deflection depending on the plasma density and frequency of the light. At the same time, the dust field affects the spacetime geometry due to its dark energy features, and factors influence how much light bends when passing near an immense object like a BH via mechanisms. When passing close to a tremendous object, the DA of a light beam can be significantly influenced by Rastall gravity. While plasma considers the DA because of its interaction with photons, Rastall gravity can alter the gravitational field, producing a different DA depending on the Rastall parameter than standard GR.

This section studies the computation of the DA in the context of a plasma medium. Plasma is another essential component that influences a gravitational lens \cite{R58, Tsupko:2021yca}. Refraction in plasma causes considered deflection. The refractive index attempts to obtain auxiliary components, which is especially important in the radio regime. Consider $v$ to be the velocity of light as it travels through hot, ionized gas to understand the impacts of plasma. The refractive index, $n(r) = c/v$ for a BH surrounded by a dust field, is given by  \cite{R41}.
\begin{equation}
n(r)=\sqrt{1-\frac{\omega^{2}_{e}}{\omega^{2}_{\infty}}A(r)},\label{x20}
\end{equation}
where $c=1$ as well as $\omega^{2}_{\infty}$ is the photon frequency measured by an observer at infinity and $\omega^{2}_{e}$ is the electron plasma frequency.
The optical metric in terms of plasma medium can be defined as
\begin{equation}
d\sigma^{2}=g^{opt}_{ij}dx^{i}dx^{j}=n^{2}\left(\frac{B(r)}{A(r)}dr^{2}
+\frac{r^{2}}{A(r)}d\phi^{2}\right).\label{x21}
\end{equation}
The optical Gaussian curvature for a BH surrounded by dust field in terms of plasma frame can be calculated as follows
\begin{eqnarray}
K&\approx& \frac{2M}{r^{3}}+\frac{(1+6k\lambda) N_d}{r^3}+\frac{12k M\lambda}{r^3}-\frac{3Q^{2}}{r^{4}}+\frac{3(2k\lambda-1)M N_d}{r^4}
+\frac{6 M Q^2}{r^{5}}+\frac{3(2-\lambda)N_d Q^2}{2r^5}\nonumber\\
&+&\frac{6M\omega^2_e}{r^3\omega^2_{\infty}}+\frac{3N_d\omega^2_e}{r^3\omega^2_{\infty}}-\frac{15\lambda k N_d\omega^2_e}{2r^3\omega^2_{\infty}}-\frac{12M N_d\omega^2_e}{r^4\omega^2_{\infty}}-\frac{5Q^2\omega^2_e}{r^4\omega^2_{\infty}}+\frac{30kM\lambda N_d\omega^2_e}{r^4\omega^2_{\infty}}+\frac{13N_d Q^2\omega^2_e}{r^5\omega^2_{\infty}}\nonumber\\
&+&\frac{26 Q^2 M\omega^2_e}{r^5\omega^2_{\infty}}
-\frac{15\lambda k N_d Q^2\omega^2_e}{r^5\omega^2_{\infty}}+O(M^{2},Q^{3},\lambda^{2}, \omega^4_e).\label{x22}
\end{eqnarray}
Using the GBT, we calculate the bending angle. To do this, we use the straight line approximation $e=\frac{b}{\sin\phi}$ at zeroth order, and the DA is obtained as
\begin{equation}
\psi_{1}\approx -\int^{\pi}_{0}\int^{\infty}_{b/\sin\phi}K\sqrt{det\vec{g}}dr d\phi.\label{x23}
\end{equation}
The DA of the BH surrounded by dust field in terms of plasma frame for the leading order terms is
computed as
\begin{eqnarray}
\psi_{1}&\approx& \frac{4M}{b}-\frac{2N_d}{b}-\frac{12N_d k\lambda}{b}+\frac{24M k\lambda}{b}-\frac{3Q^2\pi}{4b^{2}}-\frac{3M N_d\pi}{2b^2}-\frac{3M N_d k\lambda\pi}{2b^2}-\frac{4NQ^2}{3b^3}\nonumber\\
&+&\frac{8MQ^{2}}{3b^{3}}+\frac{2N_dQ^2 k \lambda}{3b^3}+\frac{6M\omega^2_e}{b\omega^2_{\infty}}+\frac{6N_d\omega^2_e}{b\omega^2_{\infty}}
-\frac{15N_d k\lambda\omega^2_e}{b\omega^2_{\infty}}-\frac{5Q^2\pi\lambda\omega^2_e}{4b^2\omega^2_{\infty}}+\frac{15M N_d k\lambda\omega^2_e}{2b^2\omega^2_{\infty}}\nonumber\\
&-&\frac{21 M N_d\pi\omega^2_e}{4b^2\omega^2_{\infty}}+\frac{52N_d Q^2\omega^2_e}{9b^3\omega^2_{\infty}}+\frac{104 M Q^2\omega^2_e}{9b^3\omega^2_{\infty}}
-\frac{20N_d Q^2k\lambda\omega^2_e}{3b^3\omega^2_{\infty}}
+O(M^{2},Q^{4},\lambda^{2}, \omega^4_e).\label{x24}
\end{eqnarray}
The DA for a BH surrounded by a dust field depends on impact parameter $b$, BH mass $M$, charge $Q$, Rastall geometric parameters $k$ and $\lambda$, field structure parameters $N_d$ as well as plasma frequencies $\omega_{\infty}$ \& $\omega_e$. Moreover, it is essential to mention here that when we ignore the plasma effects in Eq. (\ref{x24}) such that $\omega^2_{e}/\omega^2_{\infty}\rightarrow 0$, then we recover the results of Eq. (\ref{x19}) in a non-plasma medium. 

\begin{figure}[H]
\centering
\includegraphics[width=6cm,height=5.5cm]{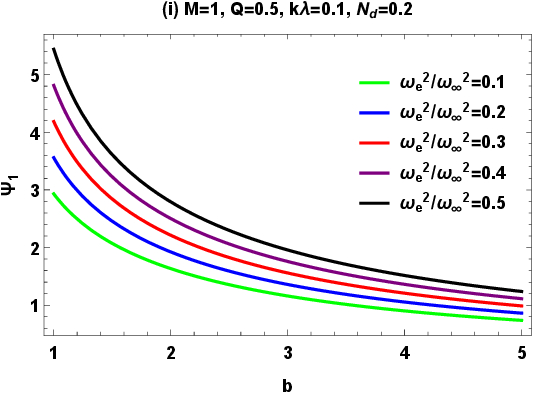}\includegraphics[width=6cm,height=5.5cm]{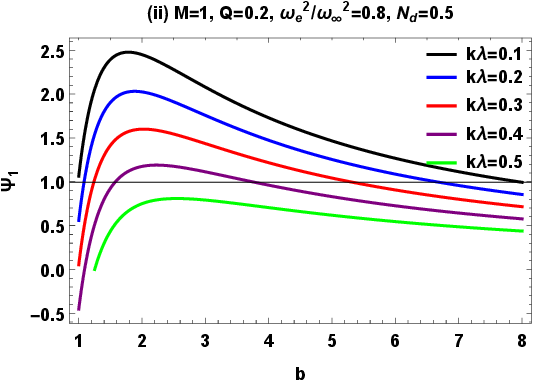}
\caption{Deflection angle $\psi_1$ w.r.t impact parameter $b$ with variations of ratio of plasma frequencies $\omega^2_e/\omega^2_{\infty}$ and Rastall parameters $k \lambda$ with fixed mass $M=1$.}\label{PD1}
\end{figure}

In Fig. \ref{PD1}: the left panel shows the variations of ratio of plasma frequencies  $\omega^2_e/\omega^2_{\infty}$ with fixed mass $M=1$, charge $Q=0.2$, Rastall parameters $k \lambda=0.1$ and dust field structure parameter $N_d=0.2$ and right panel represents the variations of Rastall parameters $k \lambda$ with fixed mass $M=1$, charge $Q=0.2$, ratio of plasma frequencies $\omega^2_e/\omega^2_{\infty}=0.8$ and dust field structure parameter $N_d=0.5$. It is evident that when the impact parameter rises, the DA for increasing plasma frequencies and Rastall parameters from a BH encircled by a dust field decreases. A significant inclination is observable at low-impact parameter values, where we observe the strongest deflection. The inverse link between impact parameter $b$ and DA $\psi_1$ is also visually apparent.

\section{Accretion disk and Shadow cast of Black holes in Rastall gravity}\label{sec5}

The photon scattering around the BH using ray tracing illustrates its physical nature. It is demonstrated that an essential variable in this ray-tracing phenomenon is the shadow cast by observer screens. To show that plasma medium influences shadow radius at massive particles. A mathematical analysis looks into the light around the properties of BHs. It is demonstrated that when photons are four-dimensional, a particular kind of dust field and non-magnetic plasma forms can encircle a shadow radius. The propagation of light can be increased or decreased by this dust field. The BH geometry significantly impacts how the shadow contributes to the overall dust field. This study examines the shadow cast by a BH surrounded by a dust field given in Eq. (\ref{x2}) and the impact of Rastall parameters on it. The core of the critical curve, also known as the apparent boundary, is represented by a BH shadow. This curve has been optimized so that light rays that are a part of it approach a bound orbit of photons asymptotically when a distant observer follows them back to the BH \cite{RALI1, Perlick:2021aok, R58, Tsukamoto:2014tja, Tsukamoto:2017fxq}.

We investigate the shadow cast by BH with a thin accretion disk. The structure that results from the gravitational influence of a central gravitating object on nearby matter, such as gas or dust, is referred to as an accretion disk \cite{v4, v5}. This matter acquires kinetic energy as it approaches the core item, creating a rapidly revolving disk around it. Depending on density and temperature, the disc may release radiation through radio waves, visible light, or X-rays. Because they make the transfer of mass and angular momentum easier, accretion disks are crucial to the formation and evolution of gravitational objects. A BH shadow appears when the lensing effect and the accretion disk agree. The BH shadow in the Newtonian case differs from that of the general-relativistic scenario primarily due to gravitational lensing, which magnifies the shadow by light bending. The shadow's contour constantly changes due to fluctuations in the light beams' orbits from the photon sphere, and the BH intrinsic properties mainly determine its size. When viewed from a distance, the shadow looks like a two-dimensional, dark disk illuminated by its uniformly brilliant surroundings. For a photon, the Hamilton-Jacobi method \cite{R62a, R62b, R62c} in the equatorial plane $\theta=\frac{\pi}{2}$ is given as
\begin{equation}
H=\frac{1}{2}\left(g^{ik}j_{i}j_{k}+\omega_{p}(r^{2})\right)=
\frac{1}{2}\left(-\frac{y^{2}_{t}}{A(r)}
+\frac{y^{2}_{r}}{B(r)}+\frac{y^{2}_{\phi}}{D(r)}
+\omega_{p}(r^{2})\right),\label{x25}
\end{equation}
where $j_i$ represents the momentum of the photon, $y_{\phi}=j_{\phi}$  is its angular momentum and $y_t=-j_t$ is the photon's energy.
The light rays are the solutions to Hamilton's equations for a derivation of the Hamiltonian Eq. (\ref{x25}) from Maxwell's equations with a two-fluid source as
\begin{equation}
\dot{y_{i}}=-\frac{\partial H}{\partial x^{i}}, ~~~~~~~~~\dot{ x^{i}}=-\frac{\partial H}{\partial y_{i}}\label{x26}.
\end{equation}
The above equations imply 
\begin{eqnarray}
\dot{y_{t}}&=&-\frac{\partial H}{\partial t}=0 ,~~~~~~~~~~
\dot{y_{\phi}}=-\frac{\partial H}{\partial \psi}=0\label{x27},\\
\dot{y_{r}}&=&-\frac{\partial H}{\partial r}=\frac{1}{2}\left(-\frac{y^{2}_{t}A'(r)}{A(r)^{2}}
+\frac{y^{2}_{r}B'(r)}{B(r)^{2}}
+\frac{y^{2}_{\phi}D'(r)}{D(r)^{2}}
-\frac{d}{dr}\omega_{p}(r^{2})\right), \label{x28}\\
\dot{t}&=&\frac{\partial H}{\partial y_{t}}=-\frac{y_{t}}{A(r)}\label{x29},\\
\dot{\phi}&=&\frac{\partial H}{\partial y_{\phi}}=-\frac{y_{\phi}}{D(r)}\label{x30},\\
\dot{r}&=&\frac{\partial H}{\partial y_{r}}=-\frac{y_{r}}{B(r)}\label{x31}.
\end{eqnarray}
In this case, a prime denotes differentiation concerning $r$, while a dot denotes differentiation to an affine parameter $\lambda$. By setting $H=0$ in Eq. (\ref{x25}), we get the result in this form
\begin{equation}
0=-\frac{y^{2}_{t}}{A(r)}
+\frac{y^{2}_{r}}{B(r)}+\frac{y^{2}_{\phi}}{D(r)}
+\omega_{p}(r^{2}).\label{x32}
\end{equation}
From Eq. (\ref{x27}), it is followed that the $y_{t}$  and $y_{\phi}$ are constants of motion. We consider $\omega_0=-j_{t}$. Given a fixed $\omega_0$ and an asymptotically flat spacetime (e. g., $A(r)\rightarrow 1$ as $r \rightarrow \infty$), the gravitational redshift formula transforms the frequency $\omega$ as measured by a static observer into a function of $r$ in the given form
\begin{equation}
\omega(r)=\frac{\omega_{0}}{\sqrt{A(r)}}.\label{x33}
\end{equation}
As a result of Eq. (\ref{x32}), a light beam with constant speed $\omega_{0}$ is limited to the area where
\begin{equation}
\frac{\omega^2_{0}}{A(r)}>\omega_{p}(r)^2.\label{x34}
\end{equation}
The constraint (\ref{x33}) states that the photon frequency $\omega(r)$ must exceed the plasma frequency $\omega_{p}(r)$ at a given place. For light to propagate through a plasma, this is always true.
To obtain the orbit equation, we use Eq. (\ref{x30}) and Eq. (\ref{x31}) and get
\begin{equation}
\frac{dr}{d\phi}=\frac{\dot{r}}{\dot{\phi}}
=\frac{D(r)y_{r}}{B(r)y_{\phi}}.\label{x34}
\end{equation}
Using $y_{r}$ from Eq. (\ref{x32}), we get the result in this form
\begin{equation}
\frac{dr}{d\phi}=\pm\frac{\sqrt{D(r)}}{\sqrt{B(r)}}
\sqrt{\frac{\omega^{2}_{0} v(r)^{2}}{y^{2}_{\phi}}-1},\label{x35}
\end{equation}
where
\begin{equation}
v(r)^{2}=\frac{D(r)}{A(r)}
\left(1-A(r)\frac{\omega_{p}^{2}}{\omega_{0}^{2}}\right).\label{x36}
\end{equation}

Since $X$ is the turning point of the trajectory, the condition $\frac{dr}{d\phi}\Big|_{X} = 0$ must hold. This equation connects $X$ to the constant of motion, $\frac{y_{\phi}}{\omega_{0}}$ as follows
\begin{equation}
v(X)^{2}=\frac{y_{\psi}^{2}}{\omega_{0}^{2}}.\label{x37}
\end{equation}
The shadow's radius is defined by the original direction of the photon's light, which is asymptotically radial towards the outer photon sphere. To calculate the shadow of the BH surrounded by a dust field, we consider a light beam projected with an angle $\gamma$ in the radial direction from the observer point $r_0$. We define $\gamma$ in the following way
\begin{equation}
\cot\gamma=\pm\frac{\sqrt{g_{rr}}}{\sqrt{g_{\phi \phi}}}\frac{dr}{d\phi}\Big|_{r=r_{0}}=
\pm\frac{\sqrt{B(r)}}{\sqrt{D(r)}}\frac{dr}{d\phi}\Big|_{r=r_{0}}.\label{x38}
\end{equation}
We rewrite the orbit equation Eq. (\ref{x35}) after attaining a minimum radius $R$ by using Eq. (\ref{x37}) in the form
\begin{equation}
\frac{dr}{d\phi}=
\pm\frac{\sqrt{D(r)}}{\sqrt{B(r)}}
\sqrt{\frac{v(r_0)^{2}}{v(X)^{2}}-1}.\label{x39}
\end{equation}
Using Eq. (\ref{x39}) into Eq. (\ref{x38}) for the angle $\gamma$, we get
\begin{equation}
\cot^{2}\gamma=
{\frac{v(r_0)^{2}}{v(X)^{2}}-1}.\label{x40}
\end{equation}
By using the identity, we obtain 
\begin{equation}
1+\cot^{2}\gamma=\frac{1}{\sin^{2}\gamma}.\label{x41}
\end{equation}
Using Eq. (\ref{x41}) into (\ref{x40}), we get the result
\begin{equation}
\sin^{2}\gamma=\frac{v(X)^{2}}{v(r_0)^{2}}.\label{x42}
\end{equation}
Light rays spiraling asymptotically towards a circular light orbit at radius $r_{ph}$ define the boundary of the shadow $\gamma$. Consequently, $X\rightarrow r_{ph}$ in Eq. (\ref{x42}) provides the angular radius of the shadow as
\begin{equation}
\sin^{2}\gamma=\frac{v(r_{ph})^{2}}{v(r_{0})^{2}},\label{x43}
\end{equation}
where $v(r_0)$ can be represented by the formula in Eq. (\ref{x35}) at $r=r_0$.
In numerous cases, it is possible to presume that the observer is situated in an area with a negligible plasma density.
Then Eq. (\ref{x35}) implies
\begin{equation}
v(r_0)^{2}=\frac{D(r_0)}{A(r_{0})},\label{x44}
\end{equation}
and Eq. (\ref{x43}) can be written as follows
\begin{equation}
\sin^{2}\gamma=
\frac{A(r_{0})D(r_{ph})}{A(r_{ph})D(r_{0})}
\left(1-\frac{A(r_{ph})\omega_{p}^{2}(r_{ph})}{\omega^{2}_{o}}\right).\label{x45}
\end{equation}
After substituting the values of the metric function into the above equation, we obtain the result in the following manner
\begin{equation}
\sin^{2}\gamma(\simeq R)=\frac{r^2_{ph}\left(2Mr_0-Q^2-r^2_0+N_d r^{\frac{1}{1-3k\lambda}}_0\right)\left(Q^2\omega^2_p-2M r_{ph}\omega^2_p+r^2_{ph}(\omega^2_P-\omega^2_0)-N_d r^{\frac{1}{1-3k\lambda}}_{ph}\omega^2_p\right)}{r^4_0\left(Q^2-2M r_{ph}+r^2_{ph}-N_d r^{\frac{1}{1-3k\lambda}}_{ph}\right)\omega^2_0}
\label{x46}.
\end{equation}
The shadow of the BH surrounded by a dust field is dependent on BH mass $M$, charge $Q$, Rastall parameters $k,~\lambda$, dust field structure parameter $N_d$, plasma frequencies $\omega_p,~\omega_0$, photon radius $r_{ph}$ and observer radius $r_0$.

\begin{table}[h]
  \centering
  \begin{tabular}{|c|c|c|c|c|c|c|}
    \hline
   $~~~~~r_{ph}~~~~~$ & ~~~~~0.5~~~~~ & 0.6 & 0.7 & 0.8 & 0.9 & 1 \\
    \hline
    $\gamma$ & 0.809198 & 1.19959 &1.68396 & 2.273 & 2.9797 & 3.82 \\
    \hline
  \end{tabular}
  \caption{Photon radius $r_{ph}$ and the shadow radius $\gamma$ for fixed $M=1=r_0,~N_d=0.5=\omega^2_{0}/\omega^2_{p}$, $k\lambda=0.1$ and $Q=0.3$.}
  \label{tab1}
\end{table}

\begin{table}[h]
  \centering
  \begin{tabular}{|c|c|c|c|c|c|c|}
    \hline
   $~~~~~r_{0}~~~~~$ & ~~~~~0.5~~~~~ & 0.6 & 0.7 & 0.8 & 0.9 & 1 \\
    \hline
    $\gamma$ & 36.6611 & 20.7166 & 12.6421 & 8.15887 & 5.49249 & 3.82 \\
    \hline
  \end{tabular}
  \caption{Observer radius $r_{0}$ and the shadow radius $\gamma$ for fixed $M=1=r_{ph},~N_d=0.5=\omega^2_{0}/\omega^2_{p}$, $k\lambda=0.1$ and $Q=0.3$.}
  \label{tab2}
\end{table}

\begin{figure}[H]
\centering
\includegraphics[width=6cm,height=6cm]{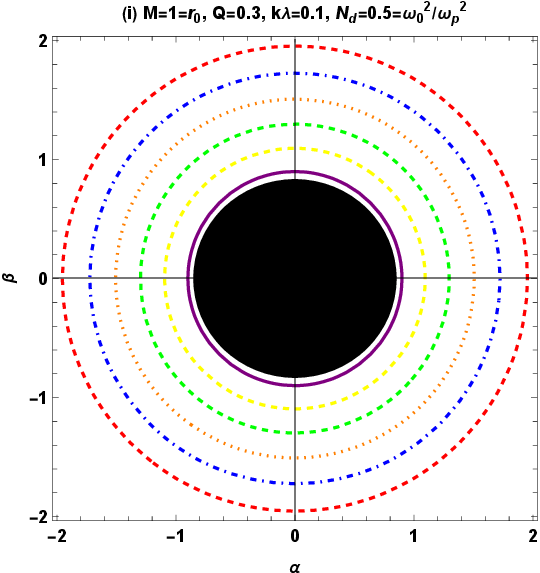}\includegraphics[width=6cm,height=6cm]{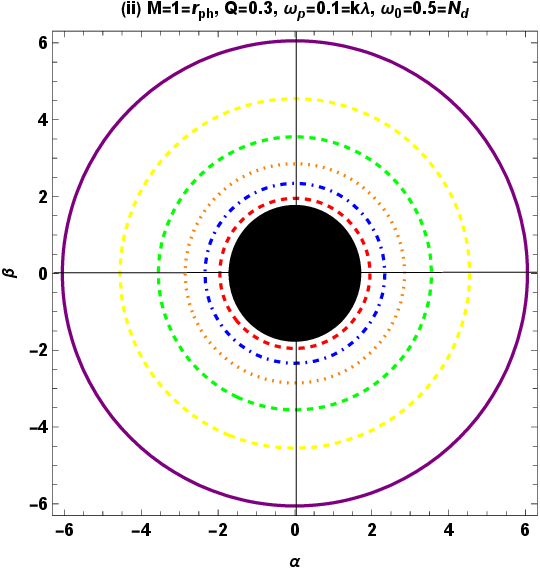}
\caption{Shadow of BH surrounded by dust field in celestial plane $(\alpha, \beta)$ for variations of photon radius $r_{ph}$ and observer radius $r_0$.}\label{CD}
\end{figure}

Fig. {\bf \ref{CD}}: the left panel shows the contour plots of shadow for variations of photon radius $r_{ph}=0.5$ (purple), $0.6$ (yellow), $0.7$ (green), $0.8$ (orange), $0.9$ (blue) and $1$ (red) with fixed values of mass $M=1$, observer radius $r_0=1$, charge $Q=0.3$, plasma frequencies $\omega_p=0.1,~\omega_0=0.5$, Rastall parameters $k\lambda=0.1$ and dust field structure parameter $N_d=0.5$. One can see that the shadow radius size increases when the photon radius rises.

The right panel gives the contour plots of shadow for different choices of radius $r_{0}=0.5$ (purple), $0.6$ (yellow), $0.7$ (green), $0.8$ (orange), $0.9$ (blue), $1$ (red) and fixed mass $M=1$, charge $Q=0.3$, photon radius $r_{ph}=1$,  plasma frequencies $\omega_p=0.1,~\omega_0=0.5$, Rastall parameters $k\lambda=0.1$ and dust field structure parameter $N_d=0.5$. It can be observed that the size of the shadow radius decreases with increasing observers radius $r_{0}$, and it is more significant than the photon radius.

\subsection{Emission energy of dust field black hole}

In BH, a dust field surrounds BH where a large concentration of small particles, known as cosmic dust, are present. These particles are often found in regions where BH forms, around dying BH, such as the dust that gives our astrophysical system its scattered light. In nature, a dust field is a collection of tiny particles that state quantum gravity through BH spacetime. Emission energy originates from quantum fluctuations in the interior of dust field BH and results from the creation and destruction of an excessive number of particles extremely near the horizon. The positive-energy particles that tunnel out of the dust field BH in the core region where Hawking radiation occurs are the primary cause of the dust field BH evaporation within a specific time frame. In this case, investigating the energy emission rate linked to the dust field BH geometry under consideration in Rastal gravity. We check the surrounding field and Rastall gravity, which influence the emission energy from a BH. The absorption cross-section frequently oscillates at a limiting constant ($\sigma_{lc}$) value with related photon radius \cite{V6, V7, V8}.
\begin{equation}
\sigma_{lc}\simeq \pi R^2,\label{E1}
\end{equation}
with $R$ denoting the dust field BH shadow radius in which the dust field BH energy emission rate is written as
\begin{equation}
\frac{d^{2}\mathbb{E}}{d\omega_{p} dt}= \frac{2\pi\sigma_{lc}\omega^{3}_{p}}{e^{\frac{\omega_{p}}{\mathbb{T}}}-1}.\label{E2}
\end{equation}
Here, Hawking temperature of dust field BH, photon emission energy, and photon frequency are denoted by $\mathbb{T},~~\mathbb{E}=\frac{d^{2}E}{d\omega_{p} dt}$ and $\omega_{p}$. By combining Eqs. (\ref{E1}) and (\ref{E2}), we can also create a new emission energy expression as
\begin{equation}
\mathbb{E}= \frac{2\pi^{3}R^2\omega^{3}_{p}}{e^{\frac{\omega_{p}}{\mathbb{T}}}-1}.\label{E3}
\end{equation}
We obtain new formulas for the shadow radius and related temperature for dust field BH. Rastall theory determines the $\mathbb{T}_{1}$ of a BH by analyzing the modified metric caused by the Rastall parameter, typically leading to a temperature shift compared to the standard Schwarzschild BH. This means that the $\mathbb{T}$ will change depending on the value of the Rastall parameter, frequently resulting in a lower temperature due to the modified gravitational effects near the event horizon. The standard $\mathbb{T}$ formula can be used, but the modified metric from Rastall gravity must be explored. The emission energy outside the dust field BH space is directly related to the BH shadow radius and photon frequency.
The BH geometry of the dust field allows us to calculate the Hawking temperature as
\begin{equation}
\mathbb{T}=\left[\frac{M}{r^{2}_{+}\pi} -\frac{Q^{2}}{r^{3}_{+}\pi
}+\frac{6k\lambda-1}{2\pi(3k\lambda-1)}
N_{d}r^{\frac{3k\lambda}{3k\lambda-1}}_{+}\right].
\end{equation}
Therefore, $\mathbb{T}$ is a dust field BH physical temperature of a function mathematically similar to the temperature in the equations of BH mechanics. Hawking examined a BH spacetime that describes gravitational collapse into a Schwarzschild BH. He evaluated a free quantum field propagating in this BH spacetime, initially in its vacuum state before the collapse, and estimated its particle content at infinity. Our computation entails taking the mass, charged, surround field, and Rastall gravity as a function corresponding to a particle state propagating in the dust field BH.

\begin{figure}[H]
\centering
\includegraphics[width=6cm,height=6cm]{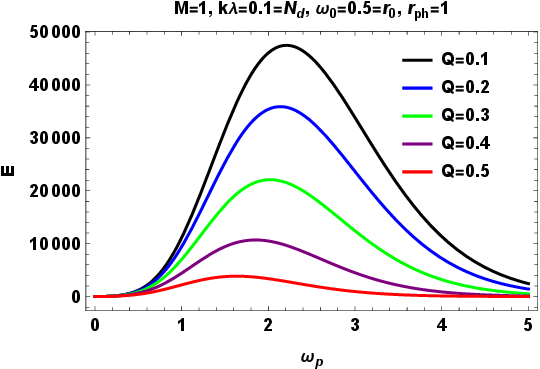}\includegraphics[width=6cm,height=6cm]{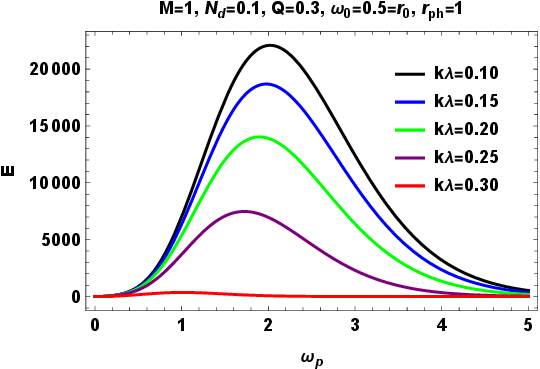}
\caption{Plots displaying the energy emission rate $E$ as a function of frequency for various values of the BH charge parameter $Q$ (Left panel) and Rastall parameter $k\lambda$
(Right panel).}\label{ED1}
\end{figure}
The left graph in Fig. \ref {ED1} shows how $E$ fluctuates with $\omega_p$ for different charge parameters $Q$ in the background of the dust field. As $Q$ grows, the peak of $E$ falls dramatically.
Most energy is produced at $Q=0.1$, decreasing as $Q$ approaches $0.5$. The curves follow a similar pattern, peaking at $\omega_p\approx2$, demonstrating a stable resonance point regardless of $Q$. For $Q=0.1$ the $E_{max}\approx50000$ and for $Q=0.5$ the $E_{max}\approx10000$. The decrease in $E$ with rising $Q$ indicates a significant inverse connection between the charge parameter $Q$ and the energy of the dust field. The parameter $Q$ affects the dust field's capacity to reach higher energies during resonance, with lower values allowing for more energy transfer.

The right graph in Fig. \ref {ED1} shows how $E$ changes as a function of $\omega_p$ for different Rastall parameter values $k\lambda$ in the background of the dust field. As $k\lambda$ grows, the maximal energy $E_{max}$ diminishes dramatically.
The resonance peak changes somewhat towards lower $\omega_p$ as $k\lambda$ increases. Higher $k\lambda$ values result in much lower total $E$ than smaller $k\lambda$. For $k\lambda=0.10$, the $E_{max}\approx20000$ and for $k\lambda=0.30$, the $E_{max}\approx2500$. The drop in $E_{max}$ indicates that raising $k\lambda$ lowers the energy density of the dust field. Greater spatial frequencies are associated with greater $k\lambda$ values, which may reduce the energy density of the dust field.

From Fig. \ref{ED1}, we conclude that $Q$ and $k\lambda$ considerably impact energy $E$, with lower values favoring more significant energy peaks. The resonance phenomena at $\omega_p\approx2$ are consistent across both graphs, indicating that it is a fundamental property of the system. Energy suppression at higher $Q$ and $k\lambda$ may be due to physical damping or dispersion in the dust field.

\section{Conclusions}\label{sec6}

In the framework of Rastall theory, we have studied generic charged BH solutions surrounded by an ideal fluid. We have examined the particular situations of BHs encircled by dust fields more closely. In the generalized theory, the physical parameters are studied by applying the weak energy condition, which signifies a positive energy density. Effective behavior of the BH surrounding the field is realized by comparing the solution in GR with the new term in the metric that arose from the Rastall theory. In the non-plasma, plasma, and perfect fluid mediums, we have discussed the Rastall solution and derived DA using the GW technique. For this purpose, the Gaussian optical curvature is determined using an optical metric, and we have examined the DA using GBT.

With the help of graphical interpretation, we showed that the impact parameter has an inverse relation with DA. Moreover, by comparing the graphical conduct of DA of our corresponding BHs with the Schwarzschild case, we concluded that BHs surrounded by dust fields have less deflection than Schwarzschild BHs. In addition, we observed strong deflection for a small value of the impact parameter in all plots of non-plasma medium. We concluded that the DA for increasing plasma frequencies and Rastall parameter values is shown to decrease via growing impact parameter from a BH surrounded by a dust field that eventually achieves an asymptotically flat form until $b\rightarrow \infty$. A notable deviation with strong deflection is observed when the impact parameter is minimal. Furthermore, the inverse link between DA and the impact parameter is also noted in the presence of a plasma medium.

At last, we investigated the shadows and emission energy of BHs defined by the Kiselev solution in the Rastall theory. The variable-separable Hamiltonian and light-ray motion equations are independent of the plasma's velocity in a pressure-less and non-magnetized plasma. We explored the perfect fluid in Rastall theory, where particle accumulation is correlated with the plasma frequency. The scenario in which plasma falls radially upon a BH from infinity is investigated in terms of a dust field encircled by a Rastall solution. We plotted the contour plots for corresponding BHs for their shadows with variations of photon radius and observer radius. We concluded that the size of the shadow radius increases with the rise in photon radius.
In contrast, the size of the shadow decreases with the increase in values of the observer's radius. The emission energy from a Rastall BH's surrounding field is the accretion disk in matter that emits intense radiation across the electromagnetic spectrum. Because of the high temperatures and extreme impact inside the disk, photons are the most noticeable emission; this process is frequently called the BH outer horizon.

The conical singularity on the symmetric axis of Rastall spacetime is responsible for the appearance of the BHs encircled by a dust field. Examining how a conical singularity affects the DA and shadow variables is interesting. Future reports on this work will be provided.

\section*{Acknowledgement}

The paper was funded by the National Natural Science Foundation of China 11975145. A. {\"O}. would like to acknowledge the contribution of the COST Action CA21106 - COSMIC WISPers in the Dark Universe: Theory, astrophysics and experiments (CosmicWISPers) and the CA22113 - Fundamental challenges in theoretical physics (THEORY-CHALLENGES).

\end{document}